# Growth and characterization of $GaN/Ga_2O_3$ Nanowire Heterostructures for Ultraviolet Optoelectronics

Edgars Butanovs[1*], Eriks Dipans[1], Martins Zubkins[1], Edvards Strods[1], Anatolijs Sarakovskis[1], Juris Purans[1], Ashutosh Kumar[2], Sergiy Khartsev[3], Martin Berg[4], Anders Hallén[3], Qin Wang[5], Joseph le Pluart[2], and Peter Ramvall[2*]

[1] Institute of Solid State Physics, University of Latvia, 8 Kengaraga Str., LV-1063 Riga, Latvia

[2] RISE Research Institutes of Sweden, Ole Römers Väg 1, SE-223 63 Lund, Sweden

[3] KTH Royal Institute of Technology, Electrum 229, SE-164 40 Kista, Sweden

[4] Hexagem AB, Ole Römers Väg 1, SE-223 63 Lund, Sweden

[5] RISE Research Institutes of Sweden, Isafjordsgatan 22, SE-164 40 Kista, Sweden

* **Corresponding author(s):** edgars.butanovs@cfi.lu.lv, peter.ramvall@ri.se

## ABSTRACT

Ultraviolet-range $GaN/\beta\text{-}Ga_2O_3$ heterostructures were fabricated and investigated in both planar and nanowire geometries using pulsed laser deposition and reactive magnetron sputtering from a liquid gallium target for $\beta\text{-}Ga_2O_3$ deposition, while both GaN nanowire arrays and planar p-type Mg-doped GaN layers were grown by metal-organic chemical vapor deposition. Precise control of film uniformity and thickness was achieved as confirmed by structural and morphology studies using X-ray diffraction, X-ray photoelectron spectroscopy, atomic force microscopy and scanning electron microscopy. Planar $n\text{-}Ga_2O_3/p\text{-}GaN$ heterojunction diodes were electrically and photoelectronically characterized, exhibiting pronounced rectifying behavior, high forward current and a visible-blind ultraviolet photoresponse under zero external bias, demonstrating intrinsic self-powered operation. Furthermore, $GaN/\beta\text{-}Ga_2O_3$ core/shell nanowire heterostructures were developed and systematically studied with a focus on morphology control and process optimization. The influence of deposition parameters on shell thickness, uniformity, and tapering was investigated, enabling improved conformality of the $\beta\text{-}Ga_2O_3$ coating on the M-plane facets of GaN nanowires. The results highlight the viability of physical vapor deposition techniques for forming $GaN/\beta\text{-}Ga_2O_3$ heterostructures and establish a pathway toward nanowire-based ultraviolet optoelectronic devices.

## I. INTRODUCTION

Ultraviolet (UV) radiation is a region of the electromagnetic spectrum commonly classified into three bands: UVA (315–400 nm), UVB (280–315 nm), and UVC (100–280 nm). Each band exhibits distinct physical and biological properties, leading to different interactions with materials, living organisms, and technological systems [1,2]. UVA radiation is widely employed in applications such as medical diagnostics, phototherapy, forensic investigations, and industrial curing processes. It is also widely used in fluorescence-based techniques and security features. UVB radiation is of considerable biological importance, most notably for its role in vitamin D synthesis in humans, and it is also utilized in clinical treatments for skin disorders including psoriasis and eczema. UVC radiation, owing to its strong germicidal efficacy, is extensively used for water purification, air treatment, and surface sterilization in healthcare and laboratory settings. In this context, semiconductor-based ultraviolet light-emitting diodes (UV LEDs) are of particular interest, as their emission wavelengths can be precisely engineered to suit specific application requirements [3,4].

Ultraviolet optoelectronic devices, especially those operating in the near-ultraviolet (UVA) region, have attracted growing attention due to their broad range of applications, including environmental monitoring, sterilization, flame detection, non-line-of-sight communication, and biomedical diagnostics [1,5,6]. In addition, solar-blind detectors sensitive to deep-ultraviolet UVC radiation are especially valuable for sterilization, water purification, and sensing applications [7–10]. Radiation in this spectral range offers inherent advantages such as low background noise and strong immunity to interference from visible and infrared light, making it a promising candidate for next-generation short-range communication technologies [7,9].

Among emerging wide-bandgap semiconductors, gallium oxide ($Ga_2O_3$) has attracted significant attention due to its outstanding intrinsic properties, including an ultra-wide bandgap (~4.5–4.9 eV), a high breakdown electric field (>$8\times10^6$ V/cm), a large Baliga's figure of merit (~3000), and the capability for growth on various heterogeneous substrates [11–22]. These characteristics have enabled $Ga_2O_3$ to be extensively explored for applications in solar-blind photodetectors, light-emitting diodes (LEDs), and gas sensors. As a photodetector material, $Ga_2O_3$ possesses a highly suitable bandgap, exhibiting an optical response peak within the solar-blind UV region and an absorption coefficient reaching up to $10^5$ $cm^{-1}$ near the absorption edge [19,23].

However, one of the major challenges facing $Ga_2O_3$ is the difficulty of achieving reliable p-type doping [24,25]. This limitation arises mainly from the inherent challenges in incorporating acceptor impurities and generating free holes within the material. Additionally, the relatively flat valence band leads to a large effective hole mass, low hole mobility, and a low diffusion constant [12,26].

In contrast, p-type gallium nitride (GaN) offers a mature and relatively well-understood platform with established doping techniques and favorable optoelectronic properties in the near-UV and visible spectral ranges. Integrating n-type $Ga_2O_3$ with p-type GaN enables the formation of a unique heterojunction that combines the advantages of both materials $Ga_2O_3$ providing deep-UV sensitivity and GaN offering efficient p-type conductivity. Such a heterostructure opens opportunities for compact, multifunctional devices capable of operating both as self-powered photodetectors and UV LEDs [13,27–31]. The p-type GaN to act as a substrate for the subsequently deposited $Ga_2O_3$ most likely has to be deposited by metal-organic chemical vapor deposition (MOCVD). However, a substantial advantage to form cost effective $Ga_2O_3$ layers/structures is to utilize deposition methods such as magnetron sputtering and pulsed laser deposition (PLD) [32].

In this work, we investigate GaN/$\beta$-$Ga_2O_3$ heterostructures with a focus on their applicability in ultraviolet optoelectronics, including self-powered photodetection. Self-powered devices, which operate without an external bias by exploiting built-in electric fields, are of particular interest for applications requiring compact and energy-efficient sensing platforms, such as environmental monitoring, space technologies, flame detection, and sterilization systems [18]. Wide-bandgap semiconductors such as $Ga_2O_3$ and GaN are especially attractive in this context due to their spectral selectivity and robustness. Furthermore, the ability to deposit $Ga_2O_3$ using scalable physical vapor deposition (PVD) techniques makes this material system promising for cost-effective device fabrication.

Two complementary GaN/$Ga_2O_3$ heterostructure configurations are explored: a planar heterojunction and a nanowire (NW)-based core/shell architecture. The planar structures are used to demonstrate heterojunction formation and evaluate basic electrical and optoelectronic properties. In parallel, the NW-based approach is investigated from a materials and structural perspective, with emphasis on the controlled deposition of $Ga_2O_3$ shells on GaN nanowires. This radial core/shell geometry offers potential advantages such as enhanced surface-to-volume ratio and improved light–matter interaction, making it a promising platform for future optoelectronic device implementations. Together, these results provide insight into both device-relevant performance and the development of nanowire heterostructures for next-generation ultraviolet technologies.

## II. EXPERIMENTAL

*MOCVD growth of GaN epilayers*

The epitaxial growth was performed in an Aixtron close-coupled showerhead (CCS) 7 x 2” wafer MOCVD tool with purified hydrogen ($H_2$), attaining a dew point less than -110 $^{o}$C, as carrier gas.

Standard precursors, such as purified ammonia ($NH_3$), trimethylgallium (TMGa), and biscyclopentadienyl magnesium ($Cp_2Mg$) were used as precursors for nitrogen, gallium, and magnesium. Epison 4 in-line gas concentration monitors were used for metal-organic precursor concentration control. The surface temperature of the wafer was optically measured and controlled by an Argus top-temperature controller (TTC). To study the development of the grown surface, the reflectance of 405 nm, 632 nm, and 951 nm light perpendicular to the surface of the sapphire substrates and GaN growth front, was measured by a LayTec EpiTT system.

As base structure for $Ga_2O_3$/GaN heterostructures Mg-doped p-GaN layers were grown on single sided polished 2″ c-plane (0001) sapphire substrates. First a standard 3-µm-thick GaN buffer layer was epitaxially grown on the sapphire substrate, followed by a 0.3-µm-thick non-intentionally doped (NID) GaN and a 0.5-µm-thick semi-insulating (SI) GaN layer. The GaN layer is made SI by increasing the carbon incorporation in the layer by adjusting the growth conditions. Finally, a 0.5-µm-thick Mg-doped GaN layer was deposited. The Mg-doped layer was grown at 995 °C with a V/III-ratio in the gas phase of 1500 at 300 mbar growth pressure. The Mg doping level targeted was $1\times10^{19}$ $cm^{-3}$ that corresponds to growth with 2.6% $Cp_2Mg$ taken as the molar flow relation between $Cp_2Mg$ and TMGa in the gas phase. No in-situ Mg-doping activation process in the MOCVD growth chamber was applied. To activate Mg doping and create free holes, rapid thermal processing (RTP) was performed in an RTP-1200-100 tool from UniTemp GmbH. The wafers were ramped to a target temperature of 900 °C at a rate of 80 °C/second in an $N_2$ ambient with an overshoot of less than 10 °C, kept at 900 °C for 5 minutes, followed by in situ cooling to about 80 °C by turning off the heater lamps. This activation procedure was found to result in a hole concentration of about $5\times10^{17}$ $cm^{-3}$ [33].

*GaN nanowire growth*

To investigate the feasibility of fabricating radial GaN/$Ga_2O_3$ core/shell nanowire (NW) heterostructures, a GaN-on-sapphire template with a p-type GaN top layer, as described above, was used as the substrate. For selective area growth (SAG) of GaN NWs, a SiN hole mask was deposited on the epitaxial GaN surface by low-pressure chemical vapor deposition (LPCVD). The mask was patterned with optical Talbot lithography to define an etch mask, followed by reactive ion etching (RIE). The holes were arranged in a hexagonal array to ensure uniform precursor supply during MOCVD growth. GaN NWs were subsequently grown in the lithographically defined holes following the procedure described in Ref. [34]. $Ga_2O_3$ shell was subsequently deposited by PLD and DC magnetron sputtering for comparison.

*Physical vapor deposition of $Ga_2O_3$*

The β-$Ga_2O_3$ layers were deposited on planar GaN/sapphire and nanowire substrates by PLD, as well as by novel direct-current (DC) magnetron sputtering from liquid target for comparison. While commonly used PLD and radiofrequency (RF) sputtering from commercial ceramic targets are the preferred options for scientific research due to their versatility, real-world manufacturing requires scalable and fast deposition solutions, such as reactive DC sputtering from metal target. Regarding the control of electrical properties, the carrier density is known to be highly sensitive to the oxygen partial pressure during PVD methods. β-$Ga_2O_3$ forms an semi-insulating layer under oxidizing conditions, whereas it exhibits n-type semiconducting behavior under reducing conditions [35–37]. However, n-type $Ga_2O_3$ films can also be fabricated by introducing Si doping into the layer [38].

PLD of β-$Ga_2O_3$ on flat substrates was performed by ablating a 99.99% purity ceramic $Ga_2O_3$ target containing 1% Si at 3 mTorr of argon/oxygen (95:5) mixture background pressure, and keeping the GaN/sapphire substrate at 600 °C. More experimental details on the PLD growth can be found in our previous work [38]. The $Ga_2O_3$:Si layer used for further planar device fabrication was about 133 nm thick as determined by ellipsometry [39] (see *Fig. S1* for more information on ellipsometry model used in this study). β-$Ga_2O_3$ thin films were also deposited by reactive pulsed-DC magnetron sputtering from a liquid gallium metal target at 150 W power in an Ar/$O_2$ atmosphere and 0.4 Pa process pressure. The substrate temperature was kept at 700 C. More technical details on the target preparation and sputtering process can be found in our previous work [40]. In this study, sputtering was used to deposit $Ga_2O_3$ around NWs, while planar reference samples were prepared for characterization needed in process development.

## III. RESULTS & DISCUSSION

*Planar n-$Ga_2O_3$:Si/p-GaN:Mg heterostructure*

X-ray diffraction (XRD) scan for $Ga_2O_3$:Si/GaN:Mg heterostructure prepared by PLD is shown in *Fig. 1(a)*, measured with Malvern Panalytical Empyrean tool with monochromatic Cu $K\alpha_1$ irradiation in ω-2θ geometry and PIXcel3D detector. The measurement confirms the crystalline nature of the layers in the heterostructure. The observed diffraction peaks at the (-201), (-402), and (-602) orientations match those expected for β-$Ga_2O_3$, verifying the presence of the thermodynamically stable β-polytype. Other Bragg peaks were attributed to the MOCVD-grown GaN and sapphire (0001) substrate.

The morphology of the as-grown p-GaN surface was characterized by atomic force microscopy

(AFM) in tapping mode by a Bruker Dimension Icon on 3.0×3.0 μm$^2$ sample surface areas, and the data analysis was done by Nanoscope software. The average ($R_a$) and root-mean-square (RMS, $R_q$) roughness were ~ 0.15 nm and ~ 0.2 nm, respectively (see *Fig. 1(b)*). AFM of the PLD-grown $Ga_2O_3$ surface revealed an average surface roughness ($R_a$) of ~ 1.5 nm and a RMS roughness ($R_q$) ~ 2.0 nm. (see *Fig. 1(c)*). AFM line scans further indicated the presence of surface crystallites protruding by ~6 nm for PLD-deposited. Comparison of PLD and magnetron sputter deposited $Ga_2O_3$ films surface morphology is shown in *Fig. S2*.

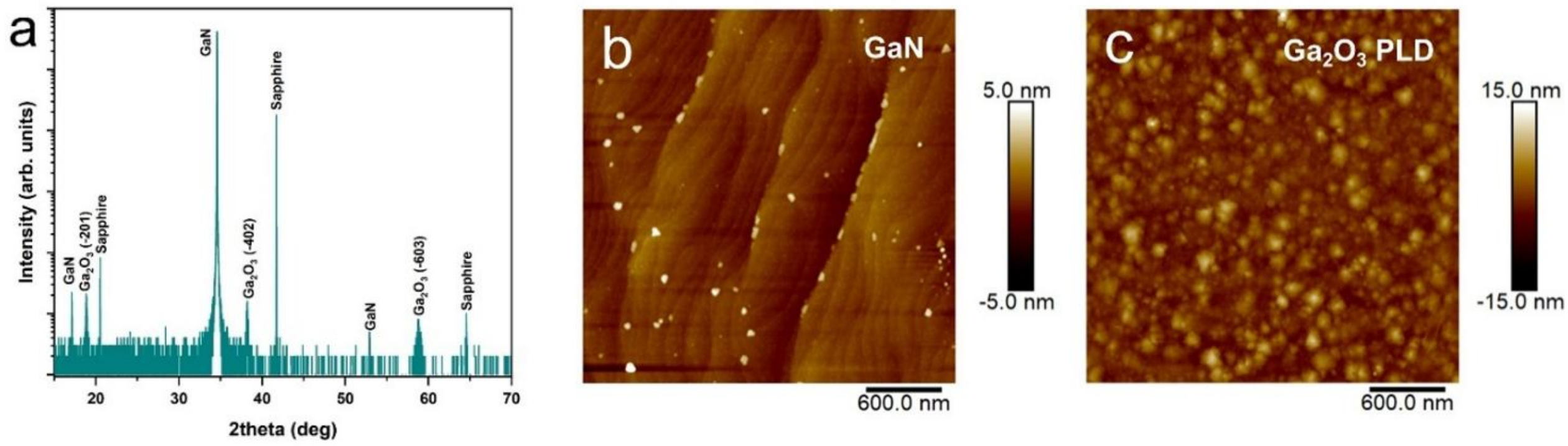

**Fig. 1.** (a) XRD diffraction pattern of $Ga_2O_3$/GaN heterostructure deposited by PLD. The diffraction peak positions corresponding to the (-201), (-402), and (-602) planes are consistent with the β-$Ga_2O_3$ phase, confirming the presence of the β-polytype. Atomic force microscopy topography images of (b) MOCVD-grown GaN epilayer surface outside the $Ga_2O_3$ deposition area, (c) PLD-grown β-$Ga_2O_3$:Si layer surface.

The chemical states of the constituent elements, the band alignment and valence band offset of the $Ga_2O_3$/GaN heterostructure were investigated in detail using X-ray photoelectron spectroscopy (XPS) and ultraviolet photoelectron spectroscopy (UPS). *Fig. 2(a, b)* presents the XPS results for a PLD-deposited doped $Ga_2O_3$:Si/GaN:Mg heterostructure. Details on the measurement realization can be found in the Supplementary Information file. In the survey spectrum of the as-prepared sample only Ga, O elements and adventitious carbon were detected. During depth profiling with $Ar^+$ ions, it was observed the XPS Ga 2p signal gradually transitions from $Ga_2O_3$ to GaN chemical state (see *Fig. 2(b)*) [13], confirming the presence of a heterointerface. The energy levels for valence band maxima (VBM) and conduction band minima for the two semiconductors are important for the p-n junction operation. To study the alignment of band edges of $Ga_2O_3$ and GaN at the junction, XPS and UPS methods are used in a similar manner as described in [41,42]. The XPS measurements provided binding energies of Ga 2p, N 1s, O 1s peaks in GaN and $Ga_2O_3$ reference samples, and GaN/$Ga_2O_3$ heterostructure, while UPS measurements yield information on VBM (see *Fig. S3* for the corresponding scans). The obtained data summarized in *Table S1* was used for energy-level diagram construction. The bandgap of GaN used in the calculation was 3.4 eV [43] while $Ga_2O_3$ – 4.9 eV [42]. The inset in *Fig. 2(b)* shows the result, displaying a Type II bandgap alignment with a relatively small shift of the conduction band edges of ΔECB = 0.1 eV, while the valence bands are separated by ΔEVB = 1.4 eV. The estimated uncertainty in the energy determination is approximately ±0.2 eV. The observation of the insignificant conduction band offset, indicating that the total bandgap difference manifests primarily as a valence band shift, is consistent with the findings reported by Wei et al. [44].

Ohmic contacts to PLD-grown n-$Ga_2O_3$:Si were formed by e-beam deposition of Ti/Au (20/100 nm) through a resist shadow mask, followed by rapid thermal annealing (RTA) at 470 °C for 1 min in $N_2$. In case of p-GaN:Mg, a stack of Ni/Au (20/100 nm) was deposited. For p-type contact formation in the planar diode heterostructure, the $Ga_2O_3$ layer was partially removed by an ICP-RIE etching step to access the underlying GaN layer. Etching was performed in Plasma-Therm APEX SLR tool for 650 seconds in 5 mTorr $BCl_3$+$Cl_2$ (10+10 sccm) gas mixture at RF=30 W and ICP = 150W power. The resulting diode structure is schematically depicted in *Fig. 2(c)*. The semiconductor layers exhibited the following carrier concentrations: for n-type $Ga_2O_3$:Si $n_e$ ~$5\times10^{18}$ $cm^{-3}$, determined from I-V characterization and an estimation of the electron mobility in equivalent samples; and for p-type GaN:Mg $n_h$ ~$5\times10^{17}$ $cm^{-3}$, measured from Hall effect.

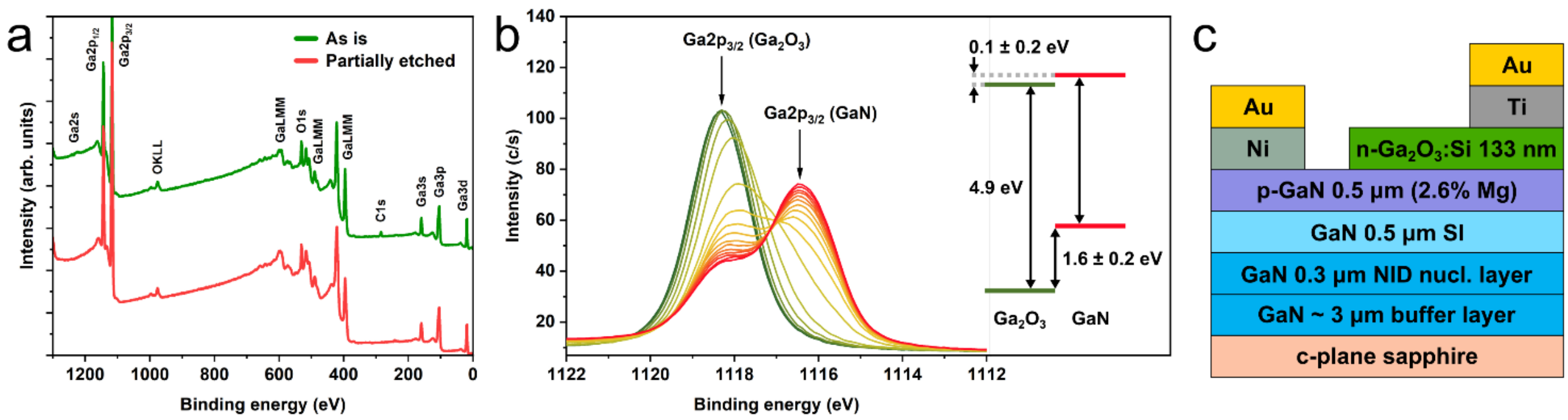

**Fig. 2.** (a) XPS survey spectrum of the $Ga_2O_3$:Si/GaN:Mg heterostructure before and after argon ion etching, indicating the constituent elements in the heterostructure. (b) High-resolution Ga $2p_{3/2}$ peak scans during depth profiling, exhibiting a gradual change from $Ga_2O_3$ to GaN chemical state. The inset shows a constructed energy level diagram for the $Ga_2O_3$:Si/GaN:Mg heterojunction. (c) A schematic of the as-fabricated planar $Ga_2O_3$:Si/GaN:Mg heterostructure.

The fabricated planar diode structure, consisting of PLD-deposited n-type $Ga_2O_3$:Si on p-type GaN:Mg was further electrically characterized by measuring its current-voltage (I-V) characteristics at both dark state (under no illumination) and when illuminated with 350 nm and 250 nm wavelength light. Biasing was performed through the Ti/Au contacts (n-side, $Ga_2O_3$) while grounding the Ni/Au contact (p-side, GaN). *Fig. 3(a)* shows the I–V characteristics in which clear rectifying behavior with higher forward-bias current (dark-state rectification ratio ~ $3x10^6$ at ±3V) can be observed. The results of the measured I-V curves under 350 nm and 250 nm wavelength light illumination exhibit self-powered operation since at 0 V applied voltage bias the diode is able to generate few nA of photocurrent. This points to suitability for applications with limited available power, for example space research.

Spectral responsivity values, defined as $R = \Delta I/(P_\lambda S)$, where $\Delta I$ is the difference between on-current and dark current, $P_\lambda$ is the light power density for the respective wavelength and S is the effective illumination area of the photodetector [45], were determined in the 250 – 400 nm range with illumination power density 280 – 3790 $\mu W/cm^2$ respectively at multiple wavelengths with 10 nm increment step, and the resulting graph in *Fig. 3(b)* shows a clear UV photoresponse and two distinct on-sets of the responsivity, most likely corresponding to GaN and $Ga_2O_3$ absorption edges at around 350 nm and 250 nm, respectively [13]. At these illumination wavelengths the corresponding responsivity values are $R_\lambda$(350 nm) = 1.06 mA/W and $R_\lambda$(250 nm) = 2.63 mA/W. The external quantum efficiency (EQE) at 0 V applied bias was calculated to be around 1.31% at 250 nm wavelength light illumination. Specific detectivity (D*) indicates the threshold for the smallest optical signal that the device can detect, which in this case is $5.4{\cdot}10^{10}$ Jones. D* was calculated as $R_\lambda/(2eI_{dark}/S)^{1/2}$, where $R_\lambda$ is the responsivity at 250 nm illumination, e is the electron charge, $I_{dark}$ is the dark current and S is the effective area of the photodetector.

Photoresponse measurements under constant cycling of illumination at negative 1 V bias voltage were carried out for about 25 seconds to test the dark current and photocurrent stability over time. In both cases (at 250 nm and 350 nm illumination, see *Fig. 3(c-d)*), dark current and photocurrent showed very little deviation over time. Time-resolved photoresponse measurements were also performed to determine rise ($t_{rise}$) and fall ($t_{fall}$) times of the photodetector, i.e. the time it takes for the photocurrent to change from 10% to 90% of its maximum value [45]. For 250 nm illumination $t_{rise}$ is 505 ms, $t_{fall}$ is 78 ms, while for 350 nm illumination $t_{rise}$ and $t_{fall}$ are 60 ms and 67 ms, respectively. The slower rise time for 250 nm wavelength could be attributed to defect trap states present in $Ga_2O_3$ layer, which has lower crystalline quality than MOCVD-grown GaN layer.

The relatively low responsivity and EQE observed here can be attributed to several factors. The relatively slow rise time under 250 nm illumination (~505 ms) indicates the presence of trap states in the $Ga_2O_3$ layer, which likely act as recombination centers and reduce carrier collection efficiency. In addition, the heterojunction is formed by ex-situ deposition, which may introduce interface states at the GaN/$Ga_2O_3$ interface, limiting the effectiveness of the built-in electric field and enhancing interfacial recombination. The current planar contact geometry may also restrict efficient carrier extraction, and improved designs such as ring-shaped contacts could significantly enhance carrier collection. Finally, the reported responsivity and EQE values correspond to zero-bias operation and therefore represents a lower bound of the device performance.

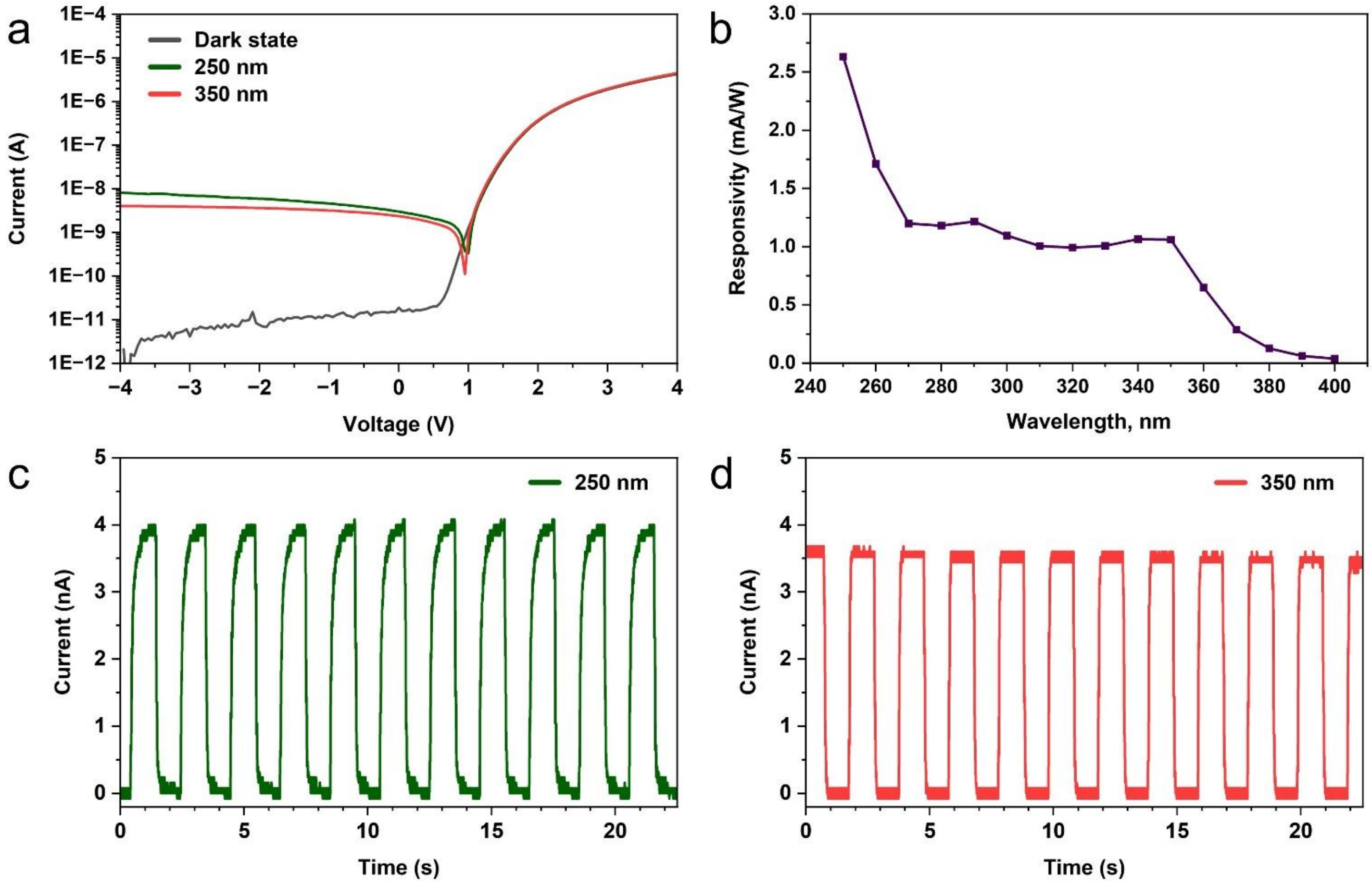

**Fig. 3.** (a) Dark-state and on-state I–V characteristics of n-$Ga_2O_3$/p-GaN diodes, showing rectifying behavior and self-powered operation. (b) Spectral responsivity curve measured at 0 V bias voltage, indicating visible-blind characteristics. (c-d) On-off curves at -1 V bias and periodic illumination of 250 nm and 350 nm wavelength light, respectively.

*GaN/$Ga_2O_3$ core/shell NW heterostructures*

The MOCVD-grown pristine GaN NWs exhibited diameters of about 120–150 nm and length of ~2 μm, as shown in *Fig. 4(a-b)*. Subsequently, $Ga_2O_3$ shells were deposited on the as-grown GaN NWs by PLD. The deposition was performed at 750 °C in 1 mTorr $O_2$, with a pulse repetition rate of 10 Hz for a total of 25 min (25,000 pulses). The resulting structures are shown in *Fig. 4(c)*. The NWs exhibit pronounced tapering, with base diameters of approximately 400 nm and a length of approximately 2 μm. Cross-sectional SEM of the substrate (not shown) indicates that 300–400 nm of $Ga_2O_3$ also accumulates on the SiN mask. Since the GaN cores maintain nearly uniform diameters along their length [34], as can be seen in *Fig. 4(b)*, the observed tapering is attributed to a $Ga_2O_3$ shell that is thinner towards the tips of the NWs. Energy-dispersive X-ray spectroscopy (EDX) of broken-off NWs (*Fig. 4(d)*) reveals a GaN NW core diameter of approximately 120–150 nm, surrounded by a ~130 nm thick polycrystalline $Ga_2O_3$ shell on the six M-plane (1-100) side facets.

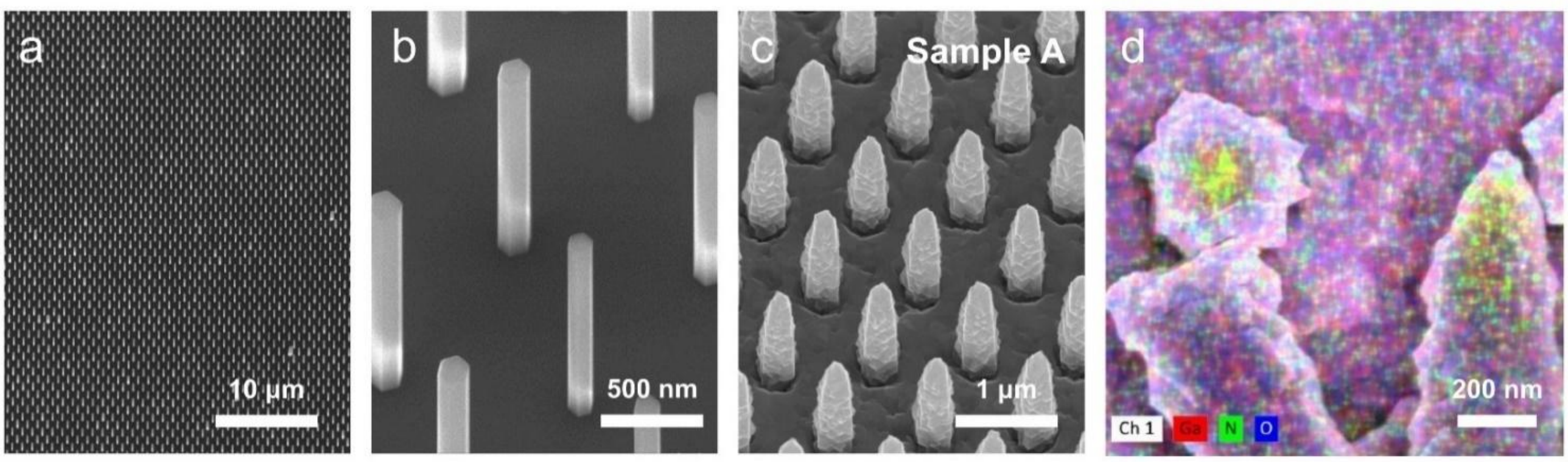

**Fig. 4.** (a-b) SEM image captured at 30° tilt angle of pristine MOCVD-grown GaN NW array at different magnifications. Occasionally malformed NWs are visible, likely caused by threading dislocations terminating within the SAG holes affecting NW nucleation. (c) SEM image captured at 30° tilt angle of GaN/$Ga_2O_3$ core/shell NWs (PLD-grown-$Ga_2O_3$, Sample A) showing a tapered $Ga_2O_3$ shell and parasitic $Ga_2O_3$ deposited on the SiN mask. (d) EDX mapping of a broken-off GaN/$Ga_2O_3$ core/shell NW. Nitrogen (green), indicating GaN, is detected predominantly at the NW core and at the NW tip, where the $Ga_2O_3$ shell is thin enough to allow electron-beam penetration.

To evaluate the effects of deposition parameters on $Ga_2O_3$ shell morphology, the deposition temperature, reactor pressure, and Ar/$O_2$ ambient mixing ratio were varied while simultaneously targeting to keep the NW length about similar. The parameter matrix is summarized in *Table 1*. The corresponding morphological results, as investigated by SEM, are shown in *Fig. 4(c)* and *Fig. 5(a-d)*.

**Table 1.** Parameters for PLD deposition of $Ga_2O_3$ on GaN NWs.

| Sample No. | Deposition Temp. | Reactor Press. | Reactor Amb. | No. of pulses |
|---|---|---|---|---|
| A | 750 °C | 1 mTorr | $O_2$ | 25,000 |
| B | 750 °C | 5 mTorr | $O_2$ | 25,000 |
| C | 550 °C | 4 mTorr | Ar/$O_2$ = 5 % | 12,500 |
| D | 550 °C | 4 mTorr | Ar | 7,500 |
| E | 550 °C | 180 mTorr | Ar/$O_2$ = 5 % | 5,000 |

*Fig. 5(a)* shows that increasing the pressure from 1 mTorr to 5 mTorr with constant temperature and ambient (Sample A and Sample B, respectively) yields slightly shorter and thicker NWs, indicating that the radial $Ga_2O_3$ deposition rate increases with higher $O_2$ pressure, whereas axial growth decreases. In *Fig. 5(b)*, the temperature was lowered to 550 °C and the reactor ambient adjusted to $Ar/O_2$ = 5 % at 4 mTorr for 12,500 laser pulses, resulting in a substantially higher deposition rate (Sample C). To determine whether this effect arises from the lower temperature or from the presence of Ar, a deposition in 100 % Ar (7,500 pulses) was conducted for Sample D under otherwise identical conditions. As seen in *Fig. 5(c)* the NW morphology shows no notable difference apart from shorter NWs due to fewer pulses, indicating that the $Ar/O_2$ ratio has only a minor influence. Finally, raising the pressure to 180 mTorr at 550 °C and $Ar/O_2$ = 5 % (5,000 pulses) for Sample E produced NWs with reversed tapering, characterized by substantially thicker top regions as seen in *Fig. 5(d)*. Furthermore, at higher pressures, the NW sidewalls also appear smoother than those grown at lower pressures. It is apparent that the reactor pressure during deposition strongly influences the resulting core/shell morphology, while deposition temperature mainly affects the deposition rate. In contrast, the $Ar/O_2$ ratio has only a minor effect.

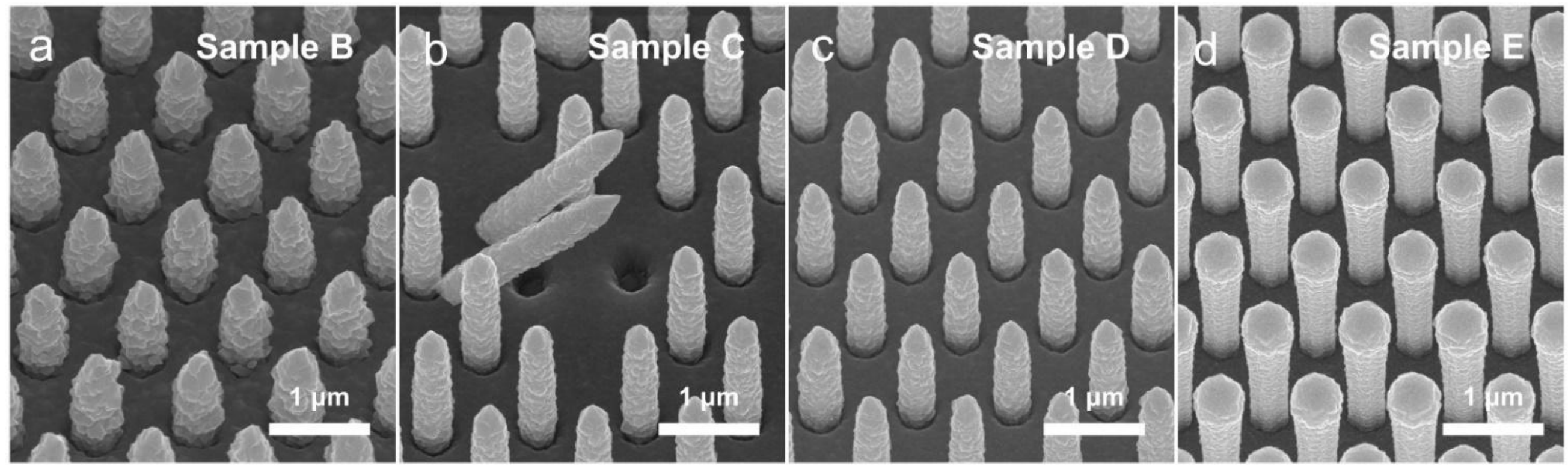

**Fig. 5.** SEM images captured at 30° tilt angle. (a) Sample B prepared at 750 °C deposition and 5 mTorr $O_2$. (b) Sample C prepared at 550 °C deposition and 4 mTorr, $Ar/O_2$ = 5 %. Several NWs appear broken, likely during handling. (c) Sample D same as (b) but with pure Ar ambient. (d) Sample E same as (b) but at 180 mTorr deposition pressure.

Deposition at elevated gas pressures is advantageous for coating surfaces oriented normal to the target due to increased scattering in the ambience. Above critical pressure, shock-wave formation produces a species distribution with narrow and wide angular components. The deposition rate is lower than in low-pressure PLD processes, where on-axis rates can exceed off-axis rates by a factor of 5–10. Bounced species may contribute significantly under these conditions, and the mechanism can change as the $Ga_2O_3$ layer on the SiN mask thickens. This may contribute to the reverse tapering observed at high pressures. Other possible factors include re-sputtering by high-energy species, differences in deposition rates on C- and M-plane facets of GaN, and variations in $Ga_2O_3$ polymorph, e.g., β-$Ga_2O_3$ on C-plane GaN versus potentially α-$Ga_2O_3$ on M-planes.

To conclude the PLD deposited GaN/$Ga_2O_3$ core/shell NW morphology investigation; varying the reactor pressure during $Ga_2O_3$ PLD growth on the M-plane facets of GaN NWs enables substantial control over the shell thickness and growth rate. By increasing the deposition pressure, the $Ga_2O_3$ deposition rate may be increased by approximately a factor of five, as evidenced by the reduction in required pulse count from 25,000 to 5,000 while maintaining NW length and $Ga_2O_3$ shell thickness (2 μm and 400 nm, respectively). Furthermore, tapering of the $Ga_2O_3$ shell, arising from differences in deposition at the NW tops and bases, can be tuned from positive to negative by adjusting the PLD reactor pressure from 1 mTorr to 180 mTorr. This observation suggests that an intermediate pressure may exist at which the $Ga_2O_3$ shell attains uniform thickness along the side of the entire GaN NW core.

An alternative, scalable PVD method for forming GaN/$Ga_2O_3$ core/shell NW heterostructures, is reactive DC magnetron sputtering from a liquid Ga target [40]. Both stoichiometric $Ga_2O_3$ deposition and reduced oxygen flow conditions, used to introduce oxygen vacancies for enhanced n-type conductivity [35–37], were investigated. Stoichiometric $Ga_2O_3$ was deposited using $O_2$ and Ar flow rates of 9.0 sccm and 30 sccm, respectively, at a chamber pressure of 3.3 mTorr. The sputtering power was 150 W, and the substrate temperature was maintained at 700 °C. Deposition was carried out in pulsed-DC mode with an on-time of 80 μs and an off-time of 20 μs.

A SEM image captured at 30° tilt angle of GaN NWs coated with stoichiometric $Ga_2O_3$ is shown in *Fig. 6(a)*. The $Ga_2O_3$ layer is partially coalesced and in the process of forming a continuous blanket layer that fully covers the GaN NWs. The cross-sectional SEM image of the same sample in *Fig. 6(b)* reveals pronounced reverse tapering of the NWs. *Fig. 6(c)* shows a cross-sectional SEM image from a position with slightly shorter NWs. The reverse tapering is even more pronounced with base and top diameters of approximately 650 nm and 900 nm, respectively. The cross-sectional SEM image also reveals that approximately 600 nm of $Ga_2O_3$ was deposited on the SiN SAG mask. From the total NW length of nearly 3 μm, the $Ga_2O_3$ deposited in the C-direction may be estimated to ~1 μm. The rightmost NW in *Fig. 6(c)* suggests the presence of such a thick $Ga_2O_3$ layer, as its upper section is cut through the thick part of the NW, but no signs of a GaN core is visible in the cross section.

An EDX investigation at 6 kV acceleration voltage of broken off GaN/$Ga_2O_3$ core/shell NWs is presented in *Fig. 6(d)*. EDX line scans, following the green line in the SEM image of a broken-off NW in the upper part in *Fig. 6(d)*. An investigation involving several NWs gave that the GaN core has a diameter of approximately 120–150 nm, and its six M-plane ($\bar{1}100$) side facets are coated with a ~250 nm-thick polycrystalline $Ga_2O_3$ shell.

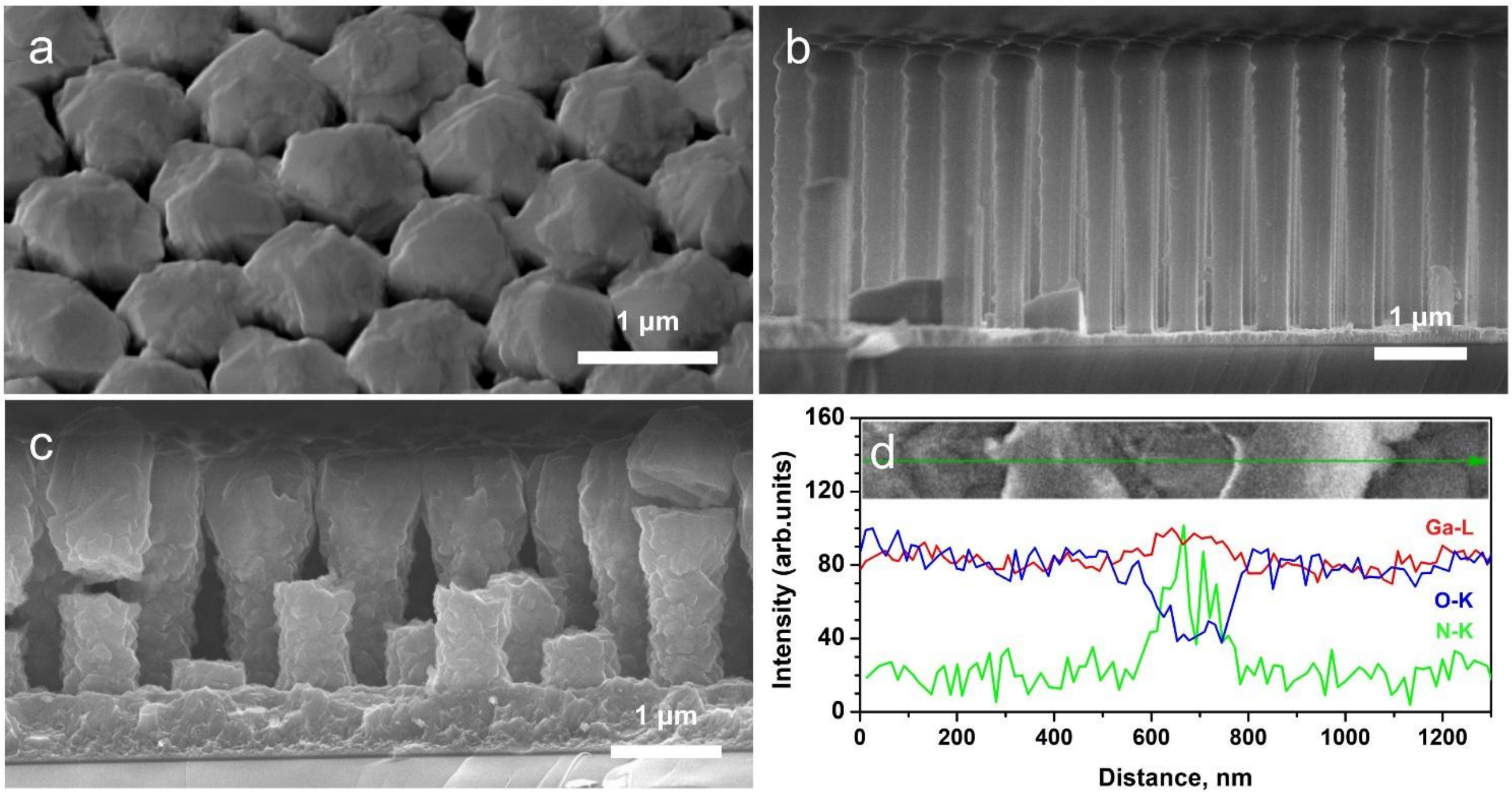

**Fig. 6.** (a) SEM image captured at 30° tilt angle of GaN/$Ga_2O_3$ core/shell NWs deposited by magnetron sputtering from a liquid Ga target at 9.0 sccm $O_2$ flow, and (b) the corresponding cross-sectional SEM image. A pronounce reverse tapering is clearly observed. (c) GaN/ $Ga_2O_3$ core/shell NWs exhibiting a nearly non-tapered diameter of ~650 nm, while the upper $Ga_2O_3$ section reaches ~900 nm. In the lower part of the SEM image, the p-type GaN layer and the 33 nm thick dark SiN selective-area growth (SAG) mask are visible. Approximately 600 nm of polycrystalline $Ga_2O_3$ is deposited on top of the SiN mask. (d) Bird's-eye-view EDX line-scan (taken along the green line in the SEM image) of a broken-off GaN/$Ga_2O_3$ core/shell NWs. Nitrogen (green), indicative of GaN, is confined to the core. The ~250 nm thick $Ga_2O_3$ shell, identified by Gallium (red) and Oxygen (blue), is polycrystalline and tightly encapsulates the GaN core.

A relatively straightforward approach to enhancing the conductivity of $Ga_2O_3$ is to carry out the deposition under oxygen-deficient conditions -[35–37]. To investigate the morphological effects of such conditions, all deposition parameters were kept constant except for the $O_2$ flow, which was reduced to 8.4 sccm. This comparatively small decrease in oxygen flow resulted in a substantial reduction in the $Ga_2O_3$ deposition rate, as shown in *Fig. 7*. Under these conditions, the diameter of the GaN/$Ga_2O_3$ NWs decreased to approximately 500 nm, corresponding to a $Ga_2O_3$ shell thickness of approximately 150–200 nm. Moreover, no evidence of $Ga_2O_3$ blanket layer formation covering the nanowires was observed.

The deposition of $Ga_2O_3$ by magnetron sputtering is highly sensitive to process parameters due to its complex phase stability and oxygen stoichiometry requirements. Even minor variations in oxygen partial pressure can drive the film from oxygen-deficient to over-oxidized regimes, significantly altering its electrical conductivity and, as observed in this work, the homogeneity of the $Ga_2O_3$ NW shell. The deposition parameter space is relatively large: sputtering power influences adatom energy and mobility, impacting crystallinity and phase formation (e.g., β-$Ga_2O_3$ versus amorphous structures). Substrate temperature further controls grain growth and defect density. Deposition chamber pressure affects mean free path and film density, while target-to-substrate distance alters deposition rate and uniformity. Because $Ga_2O_3$ properties are tightly linked to defects and microstructure, precise optimization is essential for reproducible electronic, morphological, and optoelectronic performance.

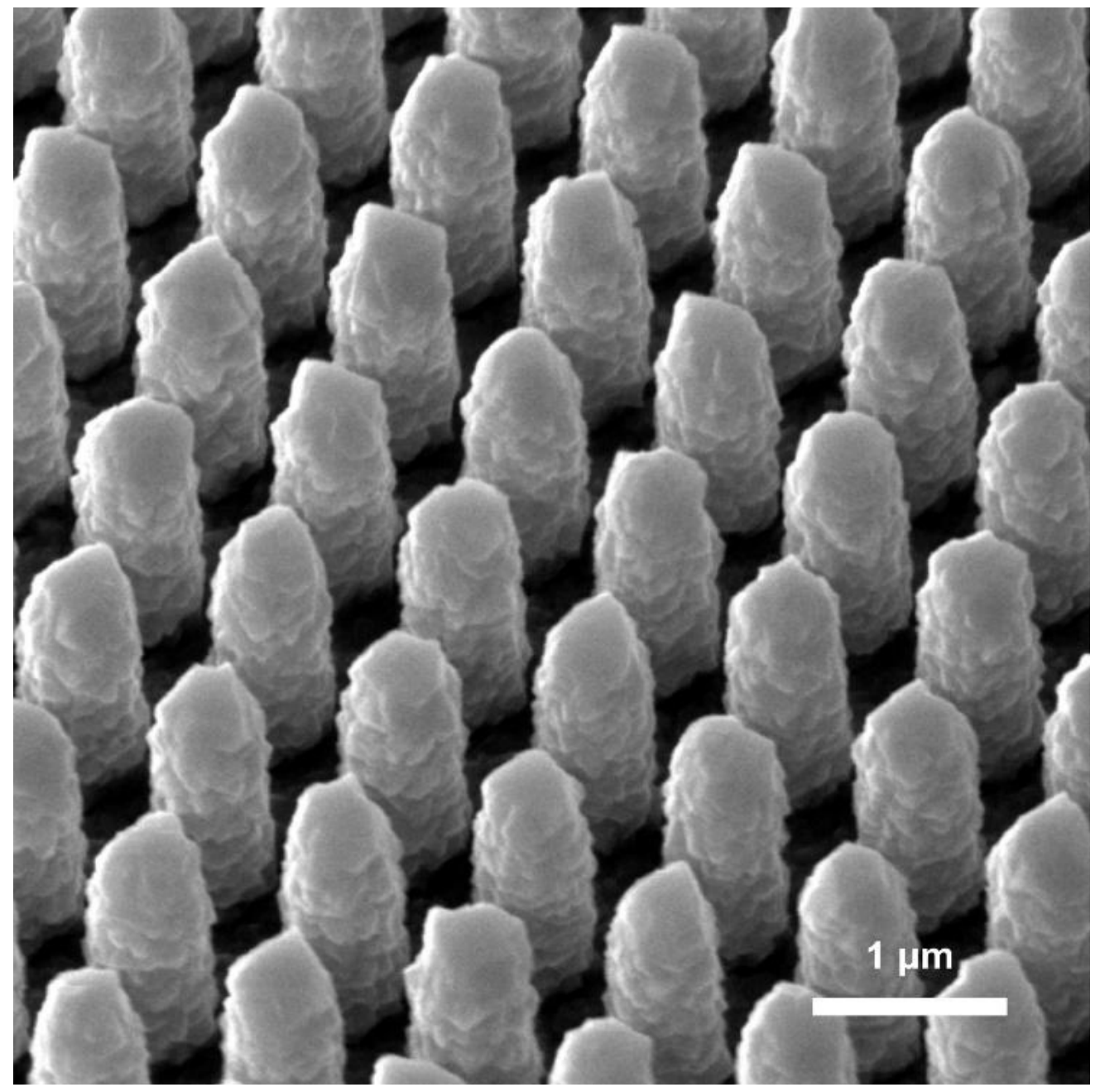

**Fig. 7.** SEM image captured at 30° tilt angle of GaN/$Ga_2O_3$ core/shell NWs deposited at 8.4 sccm $O_2$ flow. Moderate tapering is visible. The NW base diameter is ~500 nm.

## IV. CONCLUSIONS

Polycrystalline β-$Ga_2O_3$ thin films were deposited on GaN nanowires by pulsed laser deposition and magnetron sputtering from a liquid Ga target. Both methods produced $Ga_2O_3$ shells with controllable thickness, tapering behavior, and growth uniformity, enabling consistent shell formation along the entire length of the nanowires. Morphology and electrical properties were tuned through control of the Ar/$O_2$ ratio, pressure, temperature, oxygen partial pressure, and doping during deposition. While oxygen-deficient conditions can enhance electrical conductivity, they may reduce the deposition rate and adversely affect shell morphology, requiring careful optimization to balance electrical performance and structural quality.

GaN/$Ga_2O_3$ planar diode structures were fabricated and systematically characterized using scanning electron microscopy, energy-dispersive X-ray spectroscopy, ellipsometry, atomic force microscopy, X-ray diffraction, X-ray photoelectron spectroscopy, and current–voltage measurements. The devices exhibited pronounced rectifying behavior, high forward current, and a clear visible-blind ultraviolet photoresponse under zero-bias conditions, confirming self-powered operation.

GaN/$Ga_2O_3$ core/shell nanowire heterostructures were structurally investigated with a focus on morphology control. By optimizing deposition parameters, the homogeneity of the $Ga_2O_3$ shell on the M-plane facets of GaN nanowires was significantly improved, and tapering caused by nonuniform shell thickness was effectively suppressed through adjustment of deposition pressure in pulsed laser deposition and oxygen supply in magnetron sputtering. These results demonstrate the feasibility of forming well-controlled GaN/$Ga_2O_3$ nanowire heterostructures and highlight their potential for future optoelectronic devices, including vertical sensors and light-emitting diodes.

## SUPPLEMENTARY MATERIAL

See the supplementary material for supporting data on ellipsometry, AFM, and XPS analysis.

## ACKNOWLEDGMENTS

This research was funded by the European Regional Development Fund Project No. 1.1.1.3/1./24/A/020, as well as EU CAMART$^2$ project funded by European Union's Horizon 2020 Framework Programme under grant agreement No. 739508. E.D acknowledges financial support by the Project No. 1.1.1.8/1/24/I/003 "Strengthening the Research and Development Capacity of Doctoral Studies at the University of Latvia in the Fields of Smart Specialisation". The ISSP research team acknowledges the SWEB project 101087367 funded by the HORIZON-WIDERA-2022-TALENTS-01, which provided support for competence development and training.

## AUTHOR DECLARATIONS

**Conflict of Interest**

The authors have no conflicts of interest.

**Author Contributions**

**Edgars Butanovs:** Conceptualization (equal); Formal analysis (equal); Investigation (equal);

Methodology (equal); Visualization (equal); Supervision (equal); Funding acquisition (equal); Writing – original draft (equal); Project administration (equal); Writing – review & editing (equal); **Eriks Dipans**: Formal analysis (equal); Investigation (equal); Visualization (equal); Writing – original draft (equal); **Martins Zubkins**: Investigation (equal); Methodology (equal); Supervision (equal); Writing – review & editing (equal); **Edvards Strods**: Investigation (equal); Methodology (equal); **Anatolijs Sarakovskis**: Formal analysis (equal); Investigation (equal); Visualization (equal); Funding acquisition (equal); Writing – original draft (equal); Writing – review & editing (equal); **Juris Purans**: Funding acquisition (equal); Project administration (equal); Writing – review & editing (equal); **Ashutosh Kumar:** Formal analysis (equal); Investigation (equal); Methodology (equal); Visualization (equal); **Sergiy Khartsev**: Formal analysis (equal); Investigation (equal); Methodology (equal); Visualization (equal); Data Curation (equal)**; Martin Berg**: Formal analysis (equal); Investigation (equal); Methodology (equal); Visualization (equal); Data Curation (equal)**; Anders Hallén:** Conceptualization (equal); Methodology (equal); Funding acquisition (equal); **Qin Wang:** Conceptualization (equal); Funding acquisition (equal); Supervision (equal); **Joseph le Pluart:** Investigation (equal); Formal analysis (equal); Visualization (equal); Writing – review & editing (equal); **Peter Ramvall:** Conceptualization (equal); Formal analysis (equal); Investigation (equal); Methodology (equal); Visualization (equal); Funding acquisition (equal); Writing – original draft (equal); Project administration (equal); Writing – review & editing (equal); Resources (equal).

## DATA AVAILABILITY

The data that support the findings of this study are available from the corresponding author upon reasonable request.